\documentclass[11pt,a4paper]{article}
\usepackage{epsfig,amsthm,amscd,amsxtra,amsbsy}
\usepackage{amsmath,amssymb,latexsym, amsfonts}

\begin{document}

\def\DtR{\Delta t\mathcal{R}_1}
\def\fpt{\frac{\partial}{\partial t}}
\def\fpx{\frac{\partial}{\partial x}}

\newcommand{\m}[1]{\Delta x\mu_{#1+\frac{1}{2}}}
\newcommand{\vare}[1]{\Delta x\varepsilon_{#1+\frac{1}{2}}}
\newcommand{\bp}{{\mbox{}\hskip 0.2cm}}
\newcommand{\ba}{\begin{equation}}
\newcommand{\ea}{\end{equation}}
\newcommand{\bs}{\begin{split}}
\newcommand{\es}{\end{split}}

\newtheorem{theorem}{Theorem}[section]
\newtheorem{lemma}{Lemma}[section]
\newtheorem{proposition}{Proposition}[section]
\newtheorem{definition}{Definition}[section]
\newtheorem{assumption}{Assumption}[section]

\textwidth 180mm
\textheight 210mm

\baselineskip 0.6cm
\parskip 0.0cm
\parindent 0.7cm

\pagenumbering{arabic}

\title{\Large\bf Multisymplectic Geometry Method for Maxwell's Equations and Multisymplectic Scheme \thanks{
\quad Supported by the Special Funds for Major State Basic
Research Projects , G 1999, 032800}}
\author{Hongling Su\quad\quad\quad
 Mengzhao Qin\\
 E-mail: suhl@lsec.cc.ac.cn\qquad qmz@lsec.cc.ac.cn\\
 CAST (World Laboratory), \\
Institute of Computational Mathematics\\
 and Scientific/Engineering Computing, \\
Academy of Mathematics and System Sciences,\\
Chinese Academy of Sciences, Beijing, 100080, China.}
\maketitle
\begin{abstract}
In this paper we discussed the self-adjointness of the Maxwell's
equations with variable coefficients $\varepsilon$ and $\mu$.
Three different Lagrangian are attained. By the Legendre
transformation, a multisymplectic Bridge's (Hamilton) form is
obtained. Based on the multisymplectic structure, the
multisymplectic conservation law of the system is derived and a
nine-point Preissman multisymplectic scheme which preserve the
multisymplectic conservation law is given for the Maxwell's
equations in an inhomogeneous, isotropic and lossless medium. At
last a numerical example is illustrated.
\end{abstract}
\vskip 0.3cm

\section{Introduction}
Transient electromagnetic field problems are important problems in
many modern technology applications.  M. J. Gotay and J. E.
Marsden $et\quad al$\cite{Gotay} discussed the Lagrangian density
for electromagnetism on a fixed background space-time $X$ with
metric $g$, and discussed the existence of it's Euler-Lagrange
equation. In this paper we will discuss this problem from the
point of inverse variational problem. As a complete depiction of
the behavior of the electromagnetic field, Maxwell's equations
have very interesting structure.  we will discuss its Lagrangian
density and give its Multisymplectic Bridge's(Hamilton) form.
\par
The multisymplectic method is used to phrase the dynamics, which
is described by a Euler-Lagrange system, in the terms of the
$finite$-dimensional space of fields at a given event in
space-time. Some Euler-Lagarange system can be formalized in a
multisymplectic Hamilton form\cite{Marsden,Bridges1,Bridges2}
which manifest some characteristics of original systems, so we
discretize  multisymplectic Hamilton form directly in order to
simulate original system intrinsically.
\par
Determining whether a given nonlinear partial differential
equation(PDE) is an Euler-Lagrange equation of some variational
problem is fundamental work in finite-element method and nonlinear
wave theory and etc.. In this paper we will use the
straightforward result of this problem given by  Atherton and
Homsy\cite{AH} to discuss self-adjointness of Maxwell's equations
in different representations. We show that the same equations in
different forms shows different potentialness, consequently it can
be expressed as different variational problems. So we choose an
appropriate form of the Maxwell's equations to derive it's
Lagrangian and then rewrite it in a multisymplectic Hamilton
form\cite{Marsden,Bridges1,Bridges2} by using the Legendre
transformation.
\par
This paper falls into 4 parts. In Sec. 2, we obtain the necessary and sufficient conditions of self-adjointness for a
1-st order form  of the Maxwell's equations in an inhomogeneous, isotropic and lossless medium. In Sec. 3, we rewrite the
 equations in a 2-ed order form by introducing two vector potential functions. A complex multisymplectic Hamilton form and
 a multisymplectic conservation law are given as well as a ordinary conservation law. In Sec. 4, as application of
 the multisymplectic method, a multisymplectic algorithm  is constructed and numerical simulation is shown.\\
\section{Lagrangian Formalisms for Maxwell's Equations}
First we shall introduce some basic concepts in variational
problems. Let $X\subset\mathbb{R}^m$, an open subset with smooth
boundary $\partial X$. $M$ is some function space. Given a vector
differential equation with a vector of dependent variables
$u_i=u_i(x_1,x_2,\cdots,x_m)\in M,\quad i=1,2\cdot\cdot\cdot,n$,
and a vector of differential operators $N^{j}(u),\quad
j=1,2,\cdot\cdot\cdot,n$, such that \ba\label{es1.1} N(u)=0. \ea
We assume the Frech$\Acute{e}$t derivative of $N$
exist,\ba\label{es1.2} N^{\prime}_u\phi=\lim_{h\rightarrow
0}\frac{N(u+h\phi)-N(u)}{h}.\ea The arbitrary function $\phi$ now
   belongs to $M$. The Frech$\Acute{e}$t derivative derives a matrix
$N^{\prime}_u$ with  operator elements $N^{\prime}(i,j)$ which are
the derivatives of the operators $N^i$ with respect to $u_j$, so
\ba\label{es1.3} N^{\prime}_u={N^{\prime}(i,j)}_{i,j=1}^{n}. \ea
It is called as differential operator matrix of operator matrix
$N$. Then the self-adjoint test functional is given as
\ba\label{es1.4} S(\psi,\phi)=\int_X\psi
N^{\prime}_u\phi\text{d}v=\int_X\psi_i
N^{\prime}(i,j)\phi_j\text{d}v. \ea From Vainberg's theorem the
condition of self-adjointness can be given as \ba\label{es1.5}
S(\psi,\phi)=S(\phi,\psi)+\int_X\text{div}P\text{d}v, \ea if we
set the boundary condition to be free, i.e. $P$ vanishes on
$\partial X$, then \ba\label{es1.6}
N^{\prime}(i,j)=N^{\prime}(j,i)^{*}, \ea where $N^{\prime}(j,i)^*$
is denoted as the adjoint operator of $N^{\prime}(j,i)$.
\par
Surely a certain variational problem can lead to more than one
Euler-Lagrange equation and different boundary conditions,
responsibly the Lagrangian and the boundary condition are not
unique in the inverse problems of finding corresponding
functionals for a Euler-Lagrange equation, so the requirement for
free boundary conditions in Ref.\cite{AH} is too strict for some
applications such as Multi-symplectic
method\cite{Bridges1,Bridges2} which needs to verify
self-adjointness of a given system to find one of its Lagrangians
at first step of the method. For example, the wave equation such
as \ba\label{es1.6fu1} u_{tt}-u_{xx}=0, \ea according to
Ref.\cite{AH} its self-adjointness condition is\ba\label{es1.6fu2}
\frac{\partial N(u)}{\partial u_2}|_{\partial X}=0, \quad
\text{where}\quad (t,x)\in X, u_2=u_{xx},u_{tt}, \ea so it does
not satisfy this condition whatever boundary condition is given,
but it actually has Lagrangian, therefor we just set
self-adjointness condition only to be condition ($\ref{es1.6}$)
without free boundary condition.
\par
Let us consider a $k$th order functional on maps $u:
\mathbb{R}^{1,3}\longrightarrow \mathbb{R}^6$, for Maxwell's
equation, $u=[H_1,H_2,H_3,E_1,E_2,E_3]$. Here $\mathbb{R}^{1,3}$
is a (1+3)-dimensional time-space domain, denoted by coordinates
$x_j$, $j=1,\cdots,4$. $\mathbb{R}^6$ is the target space, denoted
by coordinates $y^i$, $i=1,\cdots,6$, additionally we denote the
space $T_y\mathbb{R}^6\otimes T_x^*\mathbb{R}^{1,3}$ by
$\mathfrak{R}$. The functional is defined by means of a Lagrangian
$L:$ it is a function on the space
$\mathbb{R}^{1,3}\times\mathbb{R}^6\times\mathfrak{R}$, i.e. $L$
is a smooth function defined on the bundle over
$\mathbb{R}^{1,3}\times\mathbb{R}^6$ with fiber over $(x,y)$ equal
to $\mathfrak{R}$. We denote this bundle by
$T\mathbb{R}^6\otimes_{\mathbb{R}^{1,3}\times\mathbb{R}^6}T^*\mathbb{R}^{1,3}$.
Hence the functional is defined  as \ba\label{es1.0}
\mathcal{L}=\int_{\mathbb{R}^{1,3}}L(x_j,u^i(x),u_\nu^i(x))\omega,
\ea where $\omega$ is some volume 4-form on $\mathbb{R}^{1,3}$,
$u_\nu^i=\frac{\partial u^i}{\partial x_{j_1}\cdots\partial
x_{j_{\nu}}}$, $\nu=1,2,\cdots,k$. We define Legendre
transformations as \ba\label{es1.01} \pi_i^\nu:=\frac{\partial
L}{\partial u_\nu^i}(x_j,u^i(x),u_\nu^i(x)), \ea it leads to
covariant Hamiltonian function $\mathcal{H}$ on the
multisymplectic manifold
$T^*\mathbb{R}^6\otimes_{\mathbb{R}^{1,3}\times\mathbb{R}^6}T\mathbb{R}^{1,3}$,
as \ba\label{es1.02} \mathcal{H}(x_j,u^i(x),\pi_i^\nu)=\pi_i^\nu
u_\nu^i-L(x_j,u^i(x),u_\nu^i(x))). \ea

\par
Now we consider the Maxwell's equations in an inhomogeneous,
isotropic, and lossless medium \ba\label{es1.7}
\begin{split}
\mu\frac{\partial H}{\partial t}+\nabla\times E=-K \\
-\nabla\times H+\varepsilon\frac{\partial E}{\partial t}=-J ,
\end{split}
\ea $E=[E_1,E_2,E_3]^T$ is the electric field strength,
$H=[H_1,H_2,H_3]^T$ is the magnetic field strength with
superscript $\text{T}$ denoting transpose. $\varepsilon$ is the
permittivity, $\mu$ is the permeability, both $\varepsilon$ and
$\mu$ are scalar functions of space and time variables. $J$ is the
external electric-current density, and $K$ is the external
magnetic-current density. System ($\ref{es1.7}$) can be arranged
in the form \ba\label{es1.8} \cal{G}\cdot\cal{Z}=\cal{F} , \ea
where $\mathcal{Z}=\mathcal{Z}(x,y,z,t)$ is the field vector
consisting of the components of the electric field strength $E$
and the magnetic field strength $H$ as \ba\label{es1.9}
\mathcal{Z}=[H_1, H_2, H_3, E_1, E_2, E_3]^{T} \ea simplified into
$\mathcal{Z}=[H,E]^T$, similarly
$\mathcal{F}=\mathcal{F}(x,y,z,t)$ is the source vector composed
of the external electric current source $J$ and the external
magnetic current source $K$. $\mathcal{G}$ is the operator matrix
given by \ba\label{es1.10} \mathcal{G}=
\begin{pmatrix}
{\nabla\times}
& -\varepsilon\partial_t\\
{\mu\partial_t}
& \nabla\times
\end{pmatrix},
\ea
where each element of $\mathcal{G}$ is a $3-rd$ order matrix such as
\ba\label{es1.11}
\nabla\times=
\begin{pmatrix}
{0}& -\frac{\partial}{\partial z}&\frac{\partial}{\partial y}\\
{\frac{\partial}{\partial z}}& 0 &-\frac{\partial}{\partial x}\\
{-\frac{\partial}{\partial y}}& \frac{\partial}{\partial x}& 0
\end{pmatrix}
\quad\quad\quad
\varepsilon(\mu)\partial_t=\varepsilon(\mu)\cdot
\begin{pmatrix}
{\partial_t}& 0 & 0\\
{0}& \partial_t & 0\\
{0}& 0 & \partial_t
\end{pmatrix}.
\ea Because the operator $\mathcal{G}$ is a linear operator
matrix, i.e. each element of $\mathcal{G}$ is a linear operator,
its differential operator matrix denoted as $\mathcal{G}^{\prime}$
equals $\mathcal{G}$ itself. So the adjoint operator matrix of
$\mathcal{G}^{\prime}$ is as \ba\label{es1.12} \mathcal{G}^{\prime
*}=
\begin{pmatrix}
{(\nabla\times)^\text{T}}
& \varepsilon\partial_t+\varepsilon_t\\
{-\mu\partial_t-\mu_t}
& (\nabla\times)^\text{T}
\end{pmatrix},
\ea It is obvious that the self-adjoint condition ($\ref{es1.6}$)
is that $\mathcal{G}^{\prime \text{T}}=\mathcal{G}^{\prime *}$, so
the equations have to satisfy $\varepsilon=\mu$, and the two
coefficients are independent on time variable $t$. The system in
vacuum of course satisfies
$\mathcal{G}^{\prime\text{T}}=\mathcal{G}^{\prime*}$. The systems
with such self-adjointness condition have one Lagrangian as
\ba\label{es1.12fu1}
\begin{split}
L=&\frac{1}{2}<H,\nabla\times H>+\frac{1}{2}<E,\nabla\times E>-\mu<H,E_t>-<H,J>\\
&+<E,K>,
\end{split}
\ea where $\mu=\mu(x,y,z)=\varepsilon(x,y,z)$. How to calculate
Lagrangian density form Euler-Lagrange equation is to be
introduced in the next section.
\par
Consider another form of operator matrix $\mathcal{G}$
\ba\label{es1.13}
\mathcal{G}_1=
\begin{pmatrix}
{\frac{1}{\varepsilon}\nabla\times}
& \partial_t\\
{-\partial_t}
& \frac{1}{\mu}\nabla\times
\end{pmatrix},
\ea
with the source function $\mathcal{F}$ changed in
\ba\label{es1.14}
\mathcal{F}_1=-[\frac{1}{\varepsilon}J,\quad\frac{1}{\mu}K]^{\text{T}},
\ea
the dependent variable $\mathcal{Z}$ is unchanged. But the differential operator matrix of $\mathcal{G}_1$ is not a self-adjoint operator matrix, since that
\ba\label{es1.15}
\mathcal{G}_1^{\prime *}=
\begin{pmatrix}
{A+\frac{1}{\varepsilon}(\nabla\times)^\text{T}}& -\partial_t\\
{\partial_t}& B+\frac{1}{\mu}(\nabla\times)^\text{T}
\end{pmatrix}\ne\mathcal{G}_1,
\ea
where $A$ and $B$ are two 3-rd order matrix as
\ba\label{es1.16}
A=
\begin{pmatrix}
{0}& (\frac{1}{\varepsilon})_{z} & -(\frac{1}{\varepsilon})_{y}\\
{-(\frac{1}{\varepsilon})_{z}}& 0 &(-\frac{1}{\varepsilon})_{x}\\
{(\frac{1}{\varepsilon})_{y}}& (-\frac{1}{\varepsilon})_{x}& 0
\end{pmatrix},
\quad\quad\quad
B=
\begin{pmatrix}
{0}& (\frac{1}{\mu})_{z} & -(\frac{1}{\mu})_{y}\\
{-(\frac{1}{\mu})_{z}}& 0 &(-\frac{1}{\mu})_{x}\\
{(\frac{1}{\mu})_{y}}& (-\frac{1}{\mu})_{x}& 0
\end{pmatrix},
\ea
obviously if $\mathcal{G}_1^{\prime \text{T}}=\mathcal{G}_1^{\prime *}$, $\varepsilon$ and $\mu$ are free of space variables, additionally such form excludes the condition of $\varepsilon$, $\mu=0$.
\par
So we get another Lagrangian of system ($\ref{es1.7}$)
\ba\label{es1.17}
\begin{split}
L=&\frac{1}{2\varepsilon}<H,\nabla\times H>+\frac{1}{2\mu}<E,\nabla\times E>-<H,E_t>-\frac{1}{\varepsilon}<H,J>\\
&+\frac{1}{\mu}<E,K>.
\end{split}
\ea
This Lagrangian seems equivalent to the Lagrangian ($\ref{es1.12fu1}$), but in fact the above cases actually dealt with two different systems because of different restrictions for the coefficients $\varepsilon$ and $\mu$ in their self-adjointness conditions.
\par
Both the two Lagrangian are of first order functions, and the
Legendre transformations are unnecessary in constructing their
multisymplectic Hamilton form, because the conjugate
 momentums are $H$ and $E$ multiplying some coefficients, for example in $L_2$ \\
\ba\label{es1.18}
\begin{split}
\frac{\partial L_2}{\partial\nabla\times H}=\frac{1}{2\varepsilon}H,\\
\frac{\partial L_2}{\partial\nabla\times E}=\frac{1}{2\mu}E,\\
\frac{\partial L_2}{\partial E_t}=-H.
\end{split}
\ea In fact, for linear 1-st order system which is self-adjoint,
if its coefficients satisfy some conditions, then it can be
written in multisymplectic Hamilton form directly.\\
\section{Multisymplectic Hamilton Forms for Maxwell's Equations}
Introducing two vector functions $U$ and $V$ satisfying $U_t=E$ and $V_t=H$, so the system can be rewritten in form $(2)$ with different operator matrix $\mathcal{G}$, different dependent variable $\mathcal{Z}$ and source function $\mathcal{F}$
\ba\label{es2.1}
\begin{split}
\mathcal{G}=\begin{pmatrix}
{\mu\partial_t^2}& \nabla\times\partial_t\\
{-\nabla\times\partial_t}& \varepsilon\partial_t^2
\end{pmatrix},\quad
\mathcal{Z}=[V_1, V_2, V_3, U_1, U_2, U_3]^T,\\
\mathcal{F}=-[K,J]^\text{T}.
\end{split}
\ea
The differential operator matrix $\mathcal{G}^{\prime}$ equals $\mathcal{G}$ itself. If $\varepsilon$ and $\mu$ are not dependent on spatial variables, then it is easy to demonstrate that $\mathcal{G}^{\prime}$ is self-adjoint, so the self-adjointness of the form ($\ref{es1.8}$) with sketch ($\ref{es2.1}$) is verified, and the potential functional is given by
\ba\label{es2.2}
\begin{split}
F(V,U)&=\int_\Omega\int_0^1[V,U]\mathcal{G}(\lambda\mathcal{Z})\text{d}\lambda\text{d}v\\
&=\int_\Omega(\frac{1}{2}\mu <V,V_{tt}>+\frac{1}{2}<V,\nabla\times U_t>+\frac{1}{2}\varepsilon<U,U_{tt}>\\
&-\frac{1}{2}<U,\nabla\times V_t>+UJ+VK)\text{d}v\\
&=\int_\Omega(-\frac{1}{2}\mu<V_t,V_t>-\frac{1}{2}<V_t,\nabla\times U>-\frac{1}{2}\varepsilon<U_t,U_t>\\
&+\frac{1}{2}<U_t,\nabla\times V>+UJ+VK +\text{div}P)\text{d}v,
\end{split}
\ea
where $<\cdot,\quad\cdot>$ represents the inner production of vectors. We know the Lagrangian is unique for a given Euler-Lagrange equation under  difference of a term $\text{div}P$. Here for simplicity we consider the Lagrangian as
\ba\label{es2.3}
\begin{split}
L=&\frac{1}{2}\mu<V_t,V_t>+\frac{1}{2}<V_t,\nabla\times U>\\
&+\frac{1}{2}\varepsilon<U_t,U_t>-\frac{1}{2}<U_t,\nabla\times V>-UJ-VK,
\end{split}
\ea
correspondingly the generalized conjugate momentums are
\ba\label{es2.4}
\begin{split}
P=\frac{\partial L}{\partial V_t}=\mu V_t+\frac{1}{2}\nabla\times U,\quad\quad
&\frac{\partial L}{\partial\nabla\times V}=-\frac{1}{2}U_t,\\
Q=\frac{\partial L}{\partial U_t}=\varepsilon U_t-\frac{1}{2}\nabla\times V,\quad\quad &\frac{\partial L}{\partial\nabla\times U}=\frac{1}{2}V_t.
\end{split}
\ea
We get the covariant Hamiltonian by
\ba\label{es2.5}
\begin{split}
S =&<P,V_t>+<Q,U_t>+<\frac{\partial L}{\partial\nabla\times V},\nabla\times V>\\
&+<\frac{\partial L}{\partial\nabla\times U},\nabla\times U> - L \\
=&<P,H>+<Q,E>-\frac{1}{2}\mu <H,H>-\frac{1}{2}\varepsilon <E,E>+UJ+VK ,
\end{split}
\ea
here $Z=[H,E,V,U,P,Q]^T$. So the Maxwell's equations are transformed into the following form
\ba\label{es2.7}
\begin{split}
\frac{1}{2}\nabla\times U=P-\mu H,\\
-\frac{1}{2}\nabla\times V=Q-\varepsilon E,\\
-P_t-\frac{1}{2}\nabla\times E=K,\\
-Q_t+\frac{1}{2}\nabla\times H=J,\\
V_t=H,\\
U_t=E.\\
\end{split}
\ea The above equations can be organized in the universal form
\ba\label{es2.8} \mathcal{M}Z_t+\mathcal{K}\nabla\times
Z=\nabla_{Z}S(Z), \ea where the rotation action $\nabla\times Z$
denotes $[\nabla\times H,\nabla\times E,\nabla\times
V,\nabla\times U,\nabla\times P,\nabla\times Q]^\text{T}$, and
$\mathcal{K}\nabla\times Z$ consists of three components such that
\ba\label{es2.9}\mathcal{K}\nabla\times Z=\mathcal{K}_1
Z_x+\mathcal{K}_2 Z_y+\mathcal{K}_3 Z_{z}.
\ea So
\ba\label{es2.10}
\mathcal{M}=
\begin{pmatrix}
{0}&0&0&0&0&0\\
{0}&0&0&0&0&0\\
{0}&0&0&0&-I&0\\
{0}&0&0&0&0&-I\\
{0}&0&I&0&0&0\\
{0}&0&0&I&0&0
\end{pmatrix},\quad
\mathcal{K}_i=
\begin{pmatrix}
{0}&0&0&\frac{1}{2}\mathcal{R}_i&0&0\\
{0}&0&-\frac{1}{2}\mathcal{R}_i&0&0&0\\
{0}&-\frac{1}{2}\mathcal{R}_i&0&0&0&0\\
\frac{1}{2}\mathcal{R}_i&0&0&0&0&0\\
{0}&0&0&0&0&0\\
{0}&0&0&0&0&0
\end{pmatrix},
\ea where $I$ is the identity element belonging to $R^{3\times3}$.
$\mathcal{M}$ is obviously anti-symmetric, and \ba\label{es2.11}
R_1=
\begin{pmatrix}
{0}&0&0\\
{0}&0&-1\\
{0}&1&0
\end{pmatrix}\quad\quad
R_2=
\begin{pmatrix}
{0}&0&1\\
{0}&0&0\\
{-1}&0&0
\end{pmatrix}\quad\quad
R_3=
\begin{pmatrix}
{0}&-1&0\\
{1}&0&0\\
0&0&0
\end{pmatrix},
\ea
ultimately $\mathcal{K}_i$ belongs to $R^{18\times18}$ and it is anti-symmetric. Additionally the rotation operator $\nabla\times=R_1\frac{\partial}{\partial x}+R_2\frac{\partial}{\partial y}+R_3\frac{\partial}{\partial z}$.
\par
The representation ($\ref{es2.8}$) is simplified expression in vector form, its complete extension is the multisymplectic Hamiltonian system which was first introduced by Bridges and Derks in \cite{Bridges2}. It has the multisymplectic conservative law
\ba\label{es2.12}
\frac{\partial}{\partial t}\boldsymbol{\omega}+\nabla\times\boldsymbol{\kappa}=0,
\ea
$\boldsymbol{\omega}$ and $\boldsymbol{\kappa}$ are the presymplectic forms
\ba\label{es2.13}
\boldsymbol{\omega}=\frac{1}{2}\text{d}Z\dot\wedge\mathcal{M}\text{d}Z,\quad\quad\boldsymbol{\kappa}=\frac{1}{2}\text{d}Z\dot\wedge\mathcal{K}\text{d}Z,
\ea
where $\text{d}Z=(\text{d}Z_1,\text{d}Z_2,\cdot\cdot\cdot)^{\text{T}}$, and for two $n$-order vector functions $p=(p_1,\cdot\cdot\cdot,p_n)^{\text{T}}$ and $q=(q_1,\cdot\cdot\cdot,q_n)^{\text{T}}$, notation $\dot\wedge$ is defined as
\ba\label{es2.14}
\text{d}p\dot\wedge\text{d}q=\sum\limits_{i=1}^n\text{d}p_i\wedge\text{d}q_i.
\ea

The system ($\ref{es2.8}$) also has the local multisymplectic conservation law
\ba\label{es2.15}
\frac{1}{2}\nabla\times\left[\text{d}U\dot\wedge\text{d}H-\text{d}V\dot\wedge\text{d}E\right]+\fpt(\text{d}V\dot\wedge\text{d}P+\text{d}U\dot\wedge\text{d}Q)=0.
\ea
\par
Now we return to the second case introduced in Sec. 2. We can
rewrite it in the multisymplectic Hamilton form directly
\ba\label{es2.16}
\begin{pmatrix}
0&I\\
-I&0
\end{pmatrix}\fpt\begin{pmatrix}
&H&\\
&E&
\end{pmatrix}+\begin{pmatrix}
\frac{1}{\varepsilon}\mathcal{R}_1&0\\
0&\frac{1}{\mu}\mathcal{R}_1
\end{pmatrix}\fpx\begin{pmatrix}
&H&\\
&E&
\end{pmatrix}=\begin{pmatrix}
&-J/{\varepsilon}&\\
&-K/\mu &
\end{pmatrix},
\ea
 here the covariant Hamiltonian is $S=-\frac{1}{\varepsilon}<J,H>-\frac{1}{\mu}<K,E>$. It has the multisymplectic conservation law
\ba\label{es2.17}
\nabla\times\left[\frac{1}{\varepsilon}\text{d}H\dot\wedge\text{d}H+\frac{1}{\mu}\text{d}E\dot\wedge\text{d}E\right]+\fpt(\text{d}E\dot\wedge\text{d}H)=0.
\ea\\
\section{A Multisymplectic Scheme for the Maxwell's Equations}
Maxwell's equations are simulated by many methods. Its first
numerical simulation is given by Yee\cite{Yee} in 1966, which was
developed in many more efficient algorithms. In this section, we
adopt a multisymplectic scheme to approximate the system.\par
  For simplicity, we consider $1+1$-dimensional multisymplectic form of
($\ref{es2.7}$), and suppose that the external magnetic-current
vanished, so $K=0$, and the rotation operator becomes
\ba\label{es3.1} \nabla\times Z=\mathcal{R}_1Z_x. \ea
\par
 We discretize the equations ($\ref{es2.7}$) by using midpoint scheme in both $t$ and $x$ directions,
\ba\label{es3.2}
\begin{split}
&\mathcal{R}_1\frac{U_{i+1j+\frac{1}{2}}-U_{ij+\frac{1}{2}}}{2\Delta x}=P_{i+\frac{1}{2}j+\frac{1}{2}}-\mu_{i+\frac{1}{2}}H_{i+\frac{1}{2}j+\frac{1}{2}},\\
&-\mathcal{R}_1\frac{V_{i+1j+\frac{1}{2}}-V_{ij+\frac{1}{2}}}{2\Delta x}=Q_{i+\frac{1}{2}j+\frac{1}{2}}-\varepsilon_{i+\frac{1}{2}}E_{i+\frac{1}{2}j+\frac{1}{2}},\\
&-\frac{P_{i+\frac{1}{2}j+1}-P_{i+\frac{1}{2}j}}{\Delta t}-\mathcal{R}_1\frac{E_{i+1j+\frac{1}{2}}-E_{ij+\frac{1}{2}}}{2\Delta x}=0,\\
&-\frac{Q_{i+\frac{1}{2}j+1}-Q_{i+\frac{1}{2}j}}{\Delta t}+\mathcal{R}_1\frac{H_{i+1j+\frac{1}{2}}-H_{ij+\frac{1}{2}}}{2\Delta x}=J_{i+\frac{1}{2}j+\frac{1}{2}},\\
&\frac{V_{i+\frac{1}{2}j+1}-V_{i+\frac{1}{2}j}}{\Delta t}=H_{i+\frac{1}{2}j+\frac{1}{2}},\\
&\frac{U_{i+\frac{1}{2}j+1}-U_{i+\frac{1}{2}j}}{\Delta t}=E_{i+\frac{1}{2}j+\frac{1}{2}}.
\end{split}
\ea Here the subscript $i$ and $j$ denote respectively $x$ and $t$
directions. The above scheme is called the Preissman
multisymplectic scheme which preserves the multisymplectic
conservative law ($\ref{es2.12}$) in discrete space, such as
\ba\label{es3.2fu1}
\frac{\boldsymbol{\kappa}_{i+1j+\frac{1}{2}}-\boldsymbol{\kappa}_{ij+\frac{1}{2}}}{2\Delta
x}+\frac{\boldsymbol{\omega}_{i+\frac{1}{2}j+1}-\boldsymbol{\omega}_{i+\frac{1}{2}j}}{\Delta
t}=0, \ea where \ba\label{es3.2fu2}
\begin{split}
\boldsymbol{\kappa}_{ij+\frac{1}{2}}&=\mathcal{R}_1\text{d}U_{ij+\frac{1}{2}}\dot\wedge\text{d}H_{ij+\frac{1}{2}}-\mathcal{R}_1\text{d}V_{ij+\frac{1}{2}}\dot\wedge\text{d}E_{ij+\frac{1}{2}},\\
\boldsymbol{\omega}_{i+\frac{1}{2}j}&=\text{d}V_{i+\frac{1}{2}j+1}\dot\wedge\text{d}P_{i+\frac{1}{2}j+1}+\text{d}U_{i+\frac{1}{2}j+1}\dot\wedge\text{d}Q_{i+\frac{1}{2}j+1}.
\end{split}
\ea Eliminating the auxiliary variables $V, U, P$ and $Q$ from the
Preissman scheme we get a nine-point multisymplectic integrator as
\ba\label{es3.3}
\begin{split}
&\mathcal{R}_1\frac{E_{i+2j+2}+2E_{i+2j+1}+E_{i+2j}-E_{ij+2}-2E_{ij+1}-E_{ij}}{4\Delta x}\\
=&-\mu_{i+3/2}\frac{H_{i+2j+2}+H_{i+1j+2}-H_{i+2j}-H_{i+1j}}{4\Delta t}\\
&-\mu_{i+1/2}\frac{H_{i+1j+2}+H_{ij+2}-H_{i+1j}-H_{ij}}{4\Delta t},\\
&\mathcal{R}_1\frac{H_{i+2j+2}+2H_{i+2j+1}+H_{i+2j}-H_{ij+2}-2H_{ij+1}-H_{ij}}{4\Delta x}\\
=&\varepsilon_{i+3/2}\frac{E_{i+2j+2}+E_{i+1j+2}-E_{i+2j}-E_{i+1j}}{4\Delta t}\\&+\varepsilon_{i+1/2}\frac{E_{i+1j+2}+E_{ij+2}-E_{i+1j}-E_{ij}}{4\Delta t}\\
&+(J_{i+3/2j+3/2}+J_{i+3/2j+1/2}+J_{i+1/2j+3/2}+J_{i+1/2j+1/2})/2.
\end{split}
\ea The numerical process can be depicted by the following
expression \ba\label{es3.6} AZ^{j+2}=BZ^{j+1}+CZ^{j}+\bar{J}, \ea
where \ba\label{es3.6fu1} A=\begin{pmatrix}
D&A_0&0&0\cdots&0&0&0\\
B_0 &D&A_1&0\cdots&0&0&0\\
0&B_1&D&A_2\cdots&0&0&0\\\\
\quad\quad&\ddots&\ddots&\ddots&\\\\
0&0&0&0\cdots&B_{s-1}&D&A_s\\
0&0&0&0\cdots&0&B_s&D\\
\qquad\qquad&\ddots
\end{pmatrix},
\ea \ba\label{es3.7} C=\begin{pmatrix}
-D&A_0&0&0\cdots&0&0&0\\
B_0 &-D&A_1&0\cdots&0&0&0\\
0&B_1&-D&A_2\cdots&0&0&0\\
\quad\quad&\ddots&\ddots&\ddots&\\ \\
0&0&0&0\cdots&B_{s-1}&-D&A_s\\
0&0&0&0\cdots&0&B_s&-D\\
\qquad\qquad&\ddots
\end{pmatrix},
\ea \ba\label{es3.7fu1} B=\begin{pmatrix}
-2D&\qquad&\qquad&\qquad\\
&-2D&\qquad&\qquad\\ \\
&\qquad&\ddots&\qquad\\ \\
&\qquad&\qquad&-2D
\end{pmatrix},\quad\bar{J}=\begin{pmatrix}
&a(0)+b(0)&\\
&a(1)+b(1)&\\
&\vdots& \end{pmatrix}, \ea where
$a(i)=J_{i+\frac{1}{2}j+\frac{1}{2}}+J_{i+\frac{3}{2}j+\frac{1}{2}},\quad
b(i)=J_{i+\frac{1}{2}j+\frac{3}{2}}+J_{i+\frac{3}{2}j+\frac{3}{2}}$,
and\ba\label{es3.8} D=\begin{pmatrix}
0&\DtR&\qquad&\qquad&\qquad\\
-\DtR&0&\DtR&\qquad&\qquad\\
&-\DtR&0&\DtR&\qquad\\ \quad\\
\qquad&\ddots&\ddots&\ddots&\qquad\\\quad\\
\qquad&\qquad&\qquad&-\DtR&0
\end{pmatrix}.
\ea
Set $r=1,2,\cdots$, then for every $i=1,2,\cdots,s,\cdots$, there are
\ba\label{es3.9}
A_i=\begin{pmatrix}
e_0+e_1&e_1&0&0\cdots&0&0&0\\
e_1 &e_1+e_2&e_2&0\cdots&0&0&0\\
0&e_2&e_2+e_3&e_3\cdots&0&0&0\\
&\qquad&\qquad&\ddots&\ddots&\ddots&\\ \\
0&0&0&0\cdots&e_{r-2}&e_{r-2}+e_{r-1}&e_{r-1}\\
0&0&0&0\cdots&0&e_{r-1}&e_{r-1}+e_r\\
\end{pmatrix},
\ea here $e_j=\vare{j+ir}I\in R^{3\times3},\quad j=1,2,\cdots,r$.
$B_i$ is defined similarly and just change $e_j$ in
$u_j=\m{j+ir}I\in R^{3\times3}$.
\par
To check the integrator ($\ref{es3.3}$), we consider a simple
example. We shall take $\varepsilon=\mu=1$  and $J=0$.
Furthermore, we take exact initial conditions as \ba\label{es3.4}
\begin{split}
H_1(x,0)=H_3(x,0)=0,\qquad H_2(x,0)=-\sqrt{\frac{\varepsilon}{\mu}}\sin(x),\\
E_1(x,0)=E_2(x,0)=0,\qquad E_3(x,0)=\sin(x),
\end{split}
\ea
under this initial condition we can get a exact solution as following
\ba\label{es3.5}
\begin{split}
H_1=H_3=0,\qquad H_2=-\sqrt{\frac{\varepsilon}{\mu}}\sin(x-\sqrt{\frac{1}{\varepsilon\mu}}t),\\
E_1=E_2=0,\qquad E_3=\sin(x-\sqrt{\frac{1}{\varepsilon\mu}}t),
\end{split}
\ea We compute $H_2$ in rectangles $[0,2\pi+3]\times[0,0.1]$,
$[0,2\pi+3]\times[0,1]$ and choose $\Delta x=(2\pi+3)/61$,$\Delta
t=0.01$. Fig.1 and Fig.2 shows the numerical solutions
respectively. We can see the solution is moving to $x$-direction
positively. In the rectangle $[0,2\pi+3]\times[0,10]$, the trend
of the solution is more manifest.
 
\par
 The absolute error at t-number=1000 is shown in Fig. 4 by the
'error-line', and the 'exact solution-line' is the graph of
solution $0.01\times H_2$. The former has a very regular route
which is so like the solution graph, and their vertexes are almost
at the zero points of each other. The regular similitude give a
challenge to modify the scheme which is expected in further task.

\par
For the multisymplectic form ($\ref{es2.16}$) there are many existing schemes to simulate the system and to preserve the multisymplectic conservation law ($\ref{es2.17}$), such as the usual midpoint scheme which doesn't present here. For cases of more general $\varepsilon$ and $\mu$, the discussion for their multisymplecticity is similar, so omitted here.

\end{document}